\documentclass[12pt,preprint2]{emulateapj}
\usepackage{pslatex}
\usepackage[T1]{fontenc}
\usepackage[latin1]{inputenc}
\usepackage{subfigure}
\usepackage{amsmath}
\usepackage{graphicx,graphics}
\usepackage{amssymb}
\usepackage[english]{babel}

\newcommand{\be}{\begin{equation}}
\newcommand{\ee}{\end{equation}}
\newcommand{\beqn}{\begin{eqnarray}}
\newcommand{\eeqn}{\end{eqnarray}}
\newcommand{\lap}{\lesssim}
\newcommand{\gap}{\gtrsim}

\newcommand{\msun}{M_\odot}

\newcommand{\beq}{\begin{equation}}
\newcommand{\eeq}{\end{equation}}
\newcommand{\mh}{M}

\def\ff{f_\star}
\def\vf{v}
\def\vp{v_{\parallel}}
\def\df{\langle\Delta\vp\rangle}
\def\dvp{\langle\Delta\vp^2\rangle}
\def\fs{f_\star}

\def\gap{\;\rlap{\lower 2.5pt
 \hbox{$\sim$}}\raise 1.5pt\hbox{$>$}\;}
\def\lap{\;\rlap{\lower 2.5pt
   \hbox{$\sim$}}\raise 1.5pt\hbox{$<$}\;}

\shorttitle{Black Hole Brownian Motion}
\shortauthors{Merritt et al.}
\begin{document}

\title{Brownian Motion of Black Holes in Dense Nuclei}

\author{David Merritt}
\affil{Department of Physics, Rochester Institute of Technology, 
Rochester, NY 14623, USA}

\author{Peter Berczik}
\affil{Astronomisches Rechen-Institut,
Zentrum fur Astronomie Univ. Heidelberg,
Monchhofstrasse 12-14, 69120 Heidelberg, Germany {\it and} \\
Main Astronomical Observatory, National Academy of Sciences of Ukraine,
Zabolotnoho Str., 27, 03680, Kiev, Ukraine}

\author{Frederik Laun}
\affil{Abteilung Medizinische Physik in der Radiologie,
Deutsches Krebsforschungszentrum,
Im Neuenheimer Feld 280, D-69120 Heidelberg, Germany}

\begin{abstract}

We evaluate the Brownian motion of a massive particle
(``black hole'') at the center of a galaxy using $N$-body simulations.
Our galaxy models have power-law central density
cusps like those observed at the centers of elliptical  
galaxies.
The simulations  show that the black hole  achieves
a steady-state kinetic energy that is substantially
different  than would be predicted based on the properties
of the galaxy model  in the  absence of the black hole.
The reason  appears to be that the black hole responds
to stars whose velocities have themselves  been raised
by the presence of the black hole.
Over a wide range of density slopes and black hole
masses, the black hole's mean kinetic energy is equal to
what would be predicted under the assumption that it
is in energy equipartition with stars lying within a distance
$\sim r_h/2$ from it, where $r_h$ is the black
hole's influence  radius.
The dependence of the Brownian velocity on black hole mass is 
approximately $\langle V^2\rangle\propto M_{BH}^{-1/(3-\gamma)}$
with $\gamma$ the power-law index of the stellar density
profile, $\rho\propto r^{-\gamma}$.
This is less steep than the $M_{BH}^{-1}$ dependence
predicted in a model where the effect of the black hole on the 
stellar velocities is ignored.
The influence of a stellar mass spectrum on the black hole's
Brownian motion is also evaluated and found to be consistent
with predictions from Chandrasekhar's theory.
We use these results to derive a probability function for
the mass of the Milky Way black hole based on a measurement of its
proper motion velocity.
Interesting constraints on $M_{BH}$ will require a velocity
resolution exceeding $0.5$ km s$^{-1}$.

\end{abstract}


\section{Introduction}

A massive black hole at the center of a galaxy undergoes
a random walk in momentum space as its motion is perturbed
by gravitational encounters with nearby stars.
The expected amplitude of this ``gravitational Brownian motion'' is
\begin{eqnarray}
\langle V^2\rangle \approx 3{m\over M}\sigma^2 
&\approx& (0.1\ {\rm km\ s}^{-1})^2 \left({m\over\msun}\right) 
\left({M\over 3\times 10^6\msun}\right)^{-1} \nonumber \\
& & \times \left({\sigma \over 100\ {\rm km\ s}^{-1}}\right)^2
\label{eq:1}
\end{eqnarray}
where $M$ and $m$ are the mass of the black hole and a typical
star respectively and $\sigma$ is the 1D stellar velocity 
dispersion; the brackets denote a time average.
Equation (\ref{eq:1}) follows from assuming that the
average kinetic energy of the black hole is equal to that
of the stars.
Gravitational Brownian motion is potentially interesting
for a number of reasons:
as a means to constrain black hole masses
(e.g. \cite{Backer:99,Reid:99,Reid:03});
as a mechanism for displacing black holes  from their otherwise
central locations in galaxies (e.g. \cite{BW:76});
and as a possible source of enhancement in the rate of supply of stars
to the black hole (e.g. \cite{Young:80}).

The approximate  validity  of equation (\ref{eq:1})
has been confirmed in a number of numerical studies
\citep{Miller:92,Taga:98,MM:01,CHL:02a,CHL:02b,DHM:03,
CHL:03,Makino:04}.
With two exceptions however \citep{MM:01,DHM:03},
these studies have been based on galaxy models with 
large, low-density cores, very different from the dense,
$\rho\sim r^{-\gamma}$ nuclei observed at the centers
of the galaxies known to harbor supermassive black holes.
Furthermore the black hole particle is often 
introduced into the  simulations in a non-self-consistent
way: first a black-hole-free model is constructed,
then the black hole is added, which causes the galaxy model
to evolve away from its initial state as the stellar motions
respond to  the suddenly deepened potential
(e.g. \cite{CHL:02a,CHL:02b}).
These practices complicate  the interpretation of the
numerical simulations  and make it difficult to derive
predictions about the expected behavior  of  black holes
in real nuclei.
For  example, the quantity $\sigma$ that appears in
equation (\ref{eq:1}) is well defined at the center of a
low-density galaxy containing no black hole,
but in a real galaxy,  $\sigma$ is a strong
function of radius, due both to the inhomogeneity of
the galaxy and to the presence of the black hole itself.

These considerations motivated us to undertake a new
series of numerical studies of gravitational Brownian 
motion.
Our primary goal was to evaluate  the validity of the equipartition
assumption for black holes at the centers of galaxy
models with realistically high central densities.
We carried out a set of $N$-body integrations similar in
character to those described by \cite{DHM:03},
but with a wider range of galaxy models and, in some
cases, with a spectrum of stellar masses.
All of the galaxy models have a power-law dependence
of stellar density on  radius near the center,
and the initial  stellar velocities were generated from
a distribution function that accounts for the  gravitational
force  from the  ``black hole''; hence, the  models
are in a precisely equilbrium state at time zero.

Our results can be summarized in a surprisingly simple way:
the black hole reaches a state of energy equipartition
with the stars lying at  distances $\lap r_h/2$ from
it, where $r_h$ is the black hole's gravitational influence 
radius (a precise definition of $r_h$ is given below).
By definition, stars within $r_h$ are moving largely
in response to the gravitational force from the black
hole; hence, the black hole's Brownian motion is determined
by the velocities of stars which themselves have been
raised by the presence of the black hole!
One consequence  is that the black hole's rms velocity
does  not drop as steeply with black hole mass as the $M^{-1/2}$ dependence
predicted by equation (\ref{eq:1}).
Another is that $\langle V^2\rangle$ can be substantially
different -- higher or lower -- than the value predicted by
equation (\ref{eq:1}) if $\sigma$ is measured outside of
the black hole's influence radius.

The properties  of our galaxy models and the details
of the  $N$-body integrations are presented in \S 2 and
\S3. 
The predictions of Chandrasekhar's theory of stellar
encounters are  reviewed in \S4; as we  point out there, 
Chandrasekhar's theory is essentially local in character
and does not make useful predictions about the expected 
value of $\langle V^2\rangle$ in realistic,
inhomogeneous galaxies.
Nevertheless the theory does say something definite
about the relation between $\langle V^2\rangle$ in
the single- and multi-mass cases.
Results from the $N$-body integrations are presented
in \S 5 (single stellar mass) and \S 6 (mass spectrum).
In \S 7, we use our  results to predict the expected
amplitude  of the Brownian motion for the Milky Way
black hole, and show how a measurement of
its velocity, or determination of an upper limit,
can be converted into a probability function
for its mass.

\section{Galaxy Model}

Supermassive black holes are observed at the centers of
early-type galaxies and the bulges of spiral galaxies.
The luminosity profiles of these systems
are well represented as power laws in the radius at
distances from the black hole less than $\sim r_b$, the break radius.
(So-called ``core'' galaxies also have luminosity profiles
that are power laws in the space density, but
with indices less than one.)
We adopted Dehnen's (1993) density profile for our
galaxy models:
\begin{equation}
\rho(r) = {(3-\gamma)M_{\rm gal}\over 4\pi}{a\over r^\gamma(r+a)^{4-\gamma}}
\end{equation}
where $\rho(r)$ is the stellar mass density,
$M_{\rm gal}$ is the total mass in stars, $a$ is the 
scale length (roughly speaking, the break radius)
and $\gamma$ is the logarithmic slope of the central
density cusp.

To this galaxy model was added a central point of mass $\mh$
representing the black hole.
For our purposes, it was crucial that the initial model
be in a steady state, so that any time dependence could be attributed to 
perturbations resulting from the finite-$N$ realization of the model,
and not to ill-defined departures from equilibrium.
To achieve this, the initial velocities of the stars
were generated from the unique, isotropic phase-space distribution
function that reproduces the Dehnen density law in the
smooth combined potential of the stars and the black hole particle
\citep{Tremaine:94}.
This distribution function is non-negative for $\gamma\ge 0.5$
in the presence of a central point mass;
we considered values of $\gamma$ in the range $0.5\le\gamma\le 2$.
The black hole particle was given zero velocity initially.

Unless otherwise indicated, we present our 
results in units such that $G=a=M_{\rm gal}=1$.
In these units, $\mh$ represents the ratio of the black hole mass
to the total mass in stars.
In real galactic spheroids, this ratio is approximately
$10^{-3}$, although with some scatter
\citep{MF:01,MH:03}.
We considered values of $\mh$ in the range $10^{-4}\le\mh\le 10^{-1}$.
The lower limit is fixed by the requirement that the black
hole particle be appreciably more massive than the star particles.

\begin{figure}
%\epsscale{1.0}
\plotone{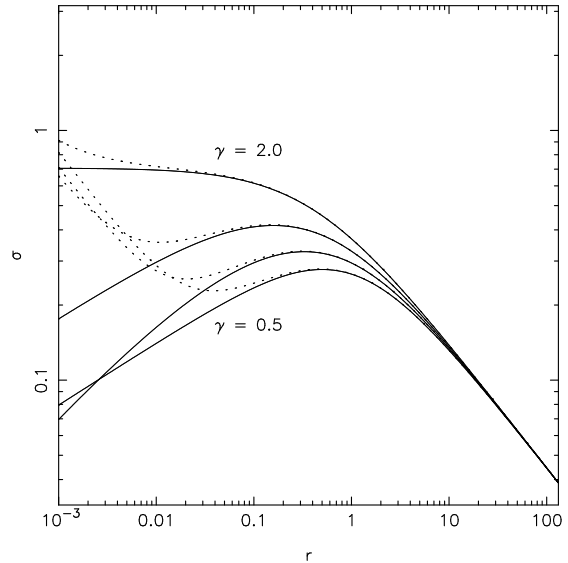}
\caption{
1D velocity dispersion profiles for Dehnen models
without central black holes (solid lines) and
with central black holes of mass $\mh=10^{-3}M_{\rm gal}$
(dashed lines).
$\gamma$ increases downward, from $\gamma=2$ to $\gamma=(1.5,1.0,0.5)$.
\label{fig:dehnen}
}
\end{figure}

Figure~\ref{fig:dehnen} shows the (1D) radial velocity dispersion 
profile $\sigma(r)$ for
Dehnen models with $\mh=0$ and $\mh=10^{-3}$.
In the absence of a black hole and for $\gamma\ne 2$, 
$\sigma(r)$ peaks at a non-zero radius;
furthermore the peak value is nearly
unaffected by the presence of a black hole of mass $10^{-3}M_{gal}$.
Since the amplitude of the Brownian motion of the black hole
is expected to scale with the stellar velocity dispersion,
it is of interest to define a characteristic value of $\sigma$
near the centers of these models.
A  natural choice is $\sigma_p(\gamma)$, the peak value
of $\sigma$ in a model without a central black hole.
We define $r_p$ as the radius at which $\sigma(r)=\sigma_p$.
Table 1 gives values of $\sigma_p$ and $r_p$
as functions of $\gamma$.

\begin{deluxetable}{lll}
\tablewidth{0pt}
\tablecaption{Dehnen model parameters}
\tablehead{
\colhead{$\gamma$} & \colhead{$\sigma_p$} & \colhead{$r_p$}
}
\startdata
0.5 & 0.2775 & 0.500 \\
1.0 & 0.3270 & 0.330 \\
1.5 & 0.4164 & 0.159 \\
2.0 & 0.7071 & 0\\
\enddata
\end{deluxetable}

Most of our $N$-body integrations were carried out on models
containing stars of equal mass.
We also carried out some integrations of models in which the stars
had a range of masses.
The stellar masses in the multi-mass integrations
were generated randomly from the mass function
\begin{equation}
n(m) dm \propto m^{-(1+\alpha)}dm
\label{eq:salpeter}
\end{equation}
with $\alpha=1.35$, the \cite{Salpeter:55} value.
Given a mean mass $\overline {m}=M_{\rm gal}/N$
with $N$ the number of particles,
the mass spectrum is determined by 
equation~(\ref{eq:salpeter}) and by $m_1/m_2$,
the ratio of smallest to largest stellar mass.
We used $m_1/m_2=0.01$.

\section{$N$-body Integrations}

All $N$-body integrations were carried out using 
a high-accuracy, direct-summation, parallel $N$-body code
\citep{BMS:05} on two supercomputers incorporating
special-purpose GRAPE \citep{FMK:05} accelerator
boards: gravitySimulator\footnote{See 
http://www.cs.rit.edu/$\sim$grapecluster/clusterInfo/grapeClusterInfo.shtml.}
and GRACE.\footnote{See http://www.ari.uni-heidelberg.de/grace.}
Particle positions were advanced using 
the fourth-order Hermite scheme, with discretized
individual time steps chosen according to the criterion
of \cite{Makino:92}, and a time-step accuracy parameter
$\eta=0.01$.
All integrations used $N=10^6$ ``star'' particles.
Inter-particle forces were softened using the standard expression,
\begin{equation}
{\bf F}_{12} = -Gm_1m_2{{\bf r}_1-{\bf r}_2\over
\left(\left|{\bf r}_1-{\bf r}_2\right|^2+\epsilon^2\right)^{3/2}};
\label{eq:plummer}
\end{equation}
the same softening length was assigned to star and to black hole
particles.

In what follows, it is argued that the black hole's Brownian
motion is determined in large part by perturbations from
stars that lie within its influence radius $r_h$, defined 
(in a slightly non-standard way) as the radius containing a
mass in stars equal to twice the black hole mass.
This means that the softening length should be small compared
with $r_h$, and in particular, small enough that the
macroscopic structure of the galaxy model does not 
change at $r \lap r_h$ due to inaccurate representation 
of the potential.
Table~2 gives values of $r_h$ for each set of $(\mh,\gamma)$
values considered here.
We verified that a softening length of $\epsilon=10^{-4}$ 
resulted in no discernible evolution in any of the models with
$\gamma\le 1.5$ even at radii $\lap 0.1r_h$.
Furthermore, as shown below (Fig.~\ref{fig:vsq1}),
our results for the $\gamma=1.5$ models using this value
of $\epsilon$ were essentially indistinguishable from
those obtained in the earlier study of Dorband et al. (2003)
who used a code with zero softening.
For the models with $\gamma=2$ and small $M$, 
some evolution at $r<r_h$ was
observed for $\epsilon=10^{-4}$ and we conservatively adopted
the smaller softening length $\epsilon=10^{-5}$ for these
models.

Even if $\epsilon$ is small enough to accurately reproduce the
small-scale structure of the galaxy models,
evolution can still occur after a sufficiently long time
as the star particles exchange energy.
In the equal-mass case, the relevant time scale is
the two-body relaxation time, 
\begin{equation}
T_r(r)={0.34 \sigma(r)^3\over\rho(r) m G^2\ln\Lambda}
\label{eq:tr}
\end{equation}
\citep{Spitzer:87},
with $\ln\Lambda$ the Coulomb logarithm.
A collisional steady state, $\rho \propto r^{-7/4}$,
is reached near the black hole in a time
$\sim T_r(r_h)$ \citep{BW:76}.
Table~2 gives estimates of $T_r(r_h)$, 
computed using $\Lambda=r_h\sigma^2(r_h)/2Gm_\star$
\citep{PMS:04}.
The short relaxation times in models with large $\gamma$
and small $\mh$ restricts how long these models can
be integrated before their structure changes.

We adopted the following integration times in model units:
\begin{eqnarray}
\gamma=0.5, && T=8 \nonumber \\
\gamma=1.0, && T=4 \nonumber \\
\gamma=1.5, && T=2 \nonumber \\
\gamma=2.0, && T=1. \nonumber
\end{eqnarray}
As Table 2 shows, the model with $(\gamma,\mh)=(2.0,3\times 10^{-4})$
has $T_r(r_h)\approx 0.16$, substantially shorter than the adopted 
integration time, and so short that it would be difficult to get good 
``statistics'' on the black hole's Brownian motion in a time $\lap T_r$.
Accordingly, we omit the results from this integration in the Table
and in the discussion that follows.

\begin{deluxetable}{llllll}
\tablewidth{0pt}
\tablecaption{Single-mass $N$-body integrations ($N=10^6$)\label{tab:two}}
\tablehead{
\colhead{$M$} & \colhead{$\gamma$} & \colhead{$r_h$} &
\colhead{$T_r(r_h)$} & \colhead{$\langle V^2\rangle$} & 
\colhead{$\eta$} 
}
\startdata

$1\times 10^{-1}$ & 0.5 & $1.11$               & $6.4\times 10^4$ & $6.72\times 10^{-6}$ & 2.91\\
                  & 1.0 & $8.09\times 10^{-1}$ & $4.3\times 10^4$ & $7.60\times 10^{-6}$ & 2.37\\
                  & 1.5 & $5.20\times 10^{-1}$ & $2.4\times 10^4$ & $1.90\times 10^{-5}$ & 3.65\\
                  & 2.0 & $2.50\times 10^{-1}$ & $9.7\times 10^3$ & $8.19\times 10^{-5}$ & 5.46\\
\hline
$3\times 10^{-2}$ & 0.5 & $4.80\times 10^{-1}$ & $1.4\times 10^4$ & $8.40\times 10^{-6}$ & 1.09\\
                  & 1.0 & $3.24\times 10^{-1}$ & $8.2\times 10^3$ & $2.04\times 10^{-5}$ & 1.91\\
                  & 1.5 & $1.81\times 10^{-1}$ & $3.6\times 10^3$ & $5.74\times 10^{-5}$ & 3.31\\
                  & 2.0 & $6.38\times 10^{-2}$ & $8.4\times 10^2$ & $2.73\times 10^{-4}$ & 5.46\\
\hline
$1\times 10^{-2}$ & 0.5 & $2.64\times 10^{-1}$ & $5.5\times 10^3$ &$2.08\times 10^{-5}$ & 0.902\\
                  & 1.0 & $1.65\times 10^{-1}$ & $2.6\times 10^3$ &$5.49\times 10^{-5}$ & 1.71\\
                  & 1.5 & $7.95\times 10^{-2}$ & $8.2\times 10^2$ &$1.24\times 10^{-4}$ & 2.39\\
                  & 2.0 & $2.04\times 10^{-2}$ & $1.0\times 10^2$ &$7.00\times 10^{-4}$ & 4.67\\
\hline
$3\times 10^{-3}$ & 0.5 & $1.48\times 10^{-1}$ & $2.4\times 10^3$ &$8.87\times 10^{-5}$ & 1.152\\
                  & 1.0 & $8.40\times 10^{-2}$ & $8.3\times 10^2$ &$9.34\times 10^{-5}$ & 0.873\\
                  & 1.5 & $3.41\times 10^{-2}$ & $1.7\times 10^2$ &$2.40\times 10^{-4}$ & 1.39\\
                  & 2.0 & $6.03\times 10^{-3}$ & $1.1\times 10^1$ &$2.13\times 10^{-3}$ & 4.26\\
\hline
$1\times 10^{-3}$ & 0.5 & $9.08\times 10^{-2}$ & $1.2\times 10^3$ &$1.03\times 10^{-4}$ & 0.445\\
                  & 1.0 & $4.68\times 10^{-2}$ & $3.1\times 10^2$ &$1.96\times 10^{-4}$ & 0.600\\
                  & 1.5 & $1.61\times 10^{-2}$ & $4.1\times 10^1$ &$5.08\times 10^{-4}$ & 0.977\\
                  & 2.0 & $2.00\times 10^{-3}$ & $1.4\times 10^0$ &$5.42\times 10^{-3}$ & 3.61\\
\hline
$3\times 10^{-4}$ & 0.5 & $5.42\times 10^{-2}$ & $6.6\times 10^2$ &$1.23\times 10^{-4}$ & 0.159\\
                  & 1.0 & $2.51\times 10^{-2}$ & $1.1\times 10^2$ &$3.58\times 10^{-4}$ & 0.335 \\
                  & 1.5 & $7.16\times 10^{-3}$ & $8.5\times 10^0$ &$1.22\times 10^{-3}$ & 0.705\\
                  & 2.0 & $6.00\times 10^{-4}$ & $1.6\times 10^{-1}$ & -- & -- \\
%\hline
%$1\times 10^{-4}$ & 0.5 & $3.43\times 10^{-2}$ & $3.9\times 10^2$ &$6.52\times 10^{-3}$ & 0.283\\
%                  & 1.0 & $1.43\times 10^{-2}$ & $4.0\times 10^1$ &$9.81\times 10^{-3}$ & 0.306\\
%                  & 1.5 & $3.43\times 10^{-3}$ & $2.1\times 10^0$    & -- & -- \\
%                  & 2.0 & $2.00\times 10^{-4}$ & $2.2\times 10^{-2}$ & -- & -- \\
\enddata
\end{deluxetable}

When evaluating the amplitude of the black hole's motion,
a possible concern is any steady component of its velocity
due to a net drift of the $N$-body system's center of mass.
The amplitude of the drift is expected to be of order
$V_{\rm drift}\approx V N^{-1/2}$, where $V$ 
is a characteristic internal (stellar) velocity.
This may be compared with the expected rms velocity 
characterizing the black hole's Brownian motion, 
which is
$\sim \sigma N^{-1/2}(\mh/M_{\rm gal})^{-1/2}$.
The latter is larger than the former by $\sim$ an order
of magnitude even for the largest black hole mass that we considered,
$\mh/M_{\rm gal}=0.1$.
We inspected the configuration-space motion of the black
hole particles in all of our integrations; 
in few if any of the cases
could we see evidence of a steady drift component
to the motion, and when it was present, it was
too small to significantly affect the estimates of 
$\langle V^2\rangle$.
Hence we ignore $V_{\rm drift}$ in what follows.

The situation is a little different in the multi-mass runs,
since a single massive particle of mass $m_\star$
can spiral into the center
in a time of order $\sim (\overline{m}/m_\star)T_r$
where $\overline{m}$ is the mean stellar mass.
The particle masses in these models were generated randomly
and a massive particle could find itself 
initially on an orbit for which the dynamical friction
time is short.
While this effect is a potential source of bias in the
multi-mass runs, the multi-mass integrations were 
shorter than the single-mass integrations, 
and we found no evidence of a significant
time dependence in the Lagrange radii, nor in the black 
hole's $\langle V^2\rangle$, that would suggest a systematic
change in the stellar distribution over the course of
the integrations.
In any case, the run-to-run variation in $\langle V^2\rangle$
due to different random realizations of the same model
were so large that they probably swamped this effect
(\S 6).

\section{Expectations from Local Theory}

Existing theories of gravitational Brownian motion
are too idealized to make clear predictions about the
amplitude of the Brownian motion in our models,
for several reasons.
First, the black hole is massive enough that
it modifies the gravitational potential as it moves.
Second, particles in tightly-bound orbits around
the black hole will increase its effective mass.
Third, the properties of the stellar background,
e.g. density and velocity dispersion,
are strong functions of distance from the black hole.
Standard encounter theory (e.g. Spitzer 1987)
approximates the stellar background as homogeneous
and time-independent and ignores the effects of
a massive particle on its surroundings.

The predictions of idealized theory are
nevertheless useful as a baseline
against which to compare the $N$-body results.
Using the Fokker-Planck equation, the steady-state
velocity distribution of a massive particle
that moves in response to perturbations
from an infinite and homogeneous distribution of
background stars is
\begin{equation}
f(V) = f_0e^{-3V^2/2\langle V^2\rangle},\ \ \ \ \langle V^2\rangle={3C\over 2A}
\end{equation}
where $\langle V^2\rangle$ is the mean square (3D) velocity
of the black hole
and $A$ and $C$ characterize the low-velocity limit of the
black hole's diffusion coefficients due to encounters with stars:
\begin{subequations}
\begin{eqnarray}
\df &=& -AV+ BV^3\ldots , \label{expand1} \\
\dvp &=& C + DV^2\ldots  \label{expand2}
\end{eqnarray}
\end{subequations}
(e.g. Merritt 2001).
The latter depend on the phase-space  number  density of stars
$\fs=\fs({\bf v}, {\bf r})$.
If $\fs$ is assumed to be isotropic in velocity space,
then
\begin{subequations}
\begin{eqnarray}
A &=& {32\over 3}\pi^2 G^2 Mmn \int_0^{\infty} {d\vf\over\vf} 
\ff(\vf) {p_{max}^2\vf^4/G^2M^2\over 1+p_{max}^2\vf^4/G^2M^2}, \\
C &=&  {16\over 3}\pi^2G^2m^2n \int_0^\infty d\vf\vf\ff(\vf)
\ln\left(1+{p_{max}^2\vf^4\over G^2M^2}\right)
\end{eqnarray}
\end{subequations}
\citep{Merritt:05}.
Here $m$ is the stellar mass, $n$ is the stellar number density,
and $p_{max}$ is the maximum effective impact parameter in 
Chandraskehar's theory.
In the special case of a Maxwellian distribution of field star velocities,
$\ff(\vf) = f_0 e^{-\vf^2/2\sigma^2}$ and
\begin{subequations}
\begin{eqnarray}
A& = & {4\sqrt{2\pi}\over 3} {G^2M mn\over\sigma^3} F(R),\ \ \ 
C = {8\sqrt{2\pi}\over 3} {G^2m^2n\over\sigma} F(R),\\
F(R) & \equiv & {1\over 2} \int_0^{\infty} dz\ e^{-z} \ln\left(1+4 R^2z^2\right), \ \ \ \ \ \ R \equiv {p_{max}\sigma^2\over GM},
\label{eq:gofr}
\end{eqnarray}
\end{subequations}
\noindent
and
\beq
\langle V^2\rangle = {3C\over 2A}=3\left({m\over M}\right)\sigma^2.
\label{eq:theory}
\eeq
The predicted $\langle V^2\rangle$ is independent of $p_{max}$ 
in this case and is equal to the ``equipartition'' value.

The velocity distribution at the center of  the Dehnen models is
not precisely Maxwellian.
While it is straightforward to derive expressions for $A$ and $C$
for non-Maxwellian $f_\star$'s \citep{Merritt:05},
a more serious problem   then presents itself.
The velocity dispersion at the  center of a Dehnen model 
is {\it zero} in the absence of a black hole (in all models 
excepting those with $\gamma=0$ or $2$);
and it is {\it infinite} when the effect of the black hole on the
stellar motions is included (Figure~\ref{fig:dehnen}).
Thus equations like (\ref{eq:theory}), naively applied, predict either
$\langle V^2\rangle=0$ or $\langle V^2\rangle=\infty$,
depending on whether or not the black  hole's influence
on the equilibrium stellar model is taken into account,
and of course neither result is physically reasonable.
The local theory fails because the Dehnen models
(like real galaxies)
are inhomogeneous, both in their density and velocity structure,
and because  the presence of the  black hole  strongly influences
even the equilibrium stellar velocity distribution in  its 
vicinity.
%In a real galaxy, the black hole will be perturbed by stars that are not 
%at the center, and the velocities of such stars are neither
%zero nor infinite.
%But Chandrasekhar's theory gives no hint about how to weight the
%contribution of these non-central stars to the black hole's
%motion.
%In such a situation, $N$-body simulations are an  indispensable
%guide.

Local theory does make one potentially useful prediction.
If the massive particle's diffusion coefficients are re-derived
for the case of a spectrum of perturber masses,
one can relate the predicted $\langle V^2\rangle$
to its value in the case of a single perturber mass.
Define the mass function such that $n(m) dm$ is the
number density of stars in mass range $m$ to $m+dm$,
and assume that the velocity distribution is the same
for all mass groups.
In the multi-mass case,
\begin{subequations}
\begin{eqnarray}
A_{mm}&=&\frac{32\pi^2}{3}G^2 M \left[\int_{m_1}^{m_2} n(m)mdm
\right] \nonumber \\
& & \int_0^\infty
\frac{d\vf}{\vf}\ff(\vf)\frac{p^2_{max}\vf^4/G^2M^2}{1+p^2_{max}\vf^4/G^2M},\\
C_{mm}&=&\frac{16\pi^2}{3}G^2 \left[\int_{m_1}^{m_2} n(m)m^2dm\right]
\nonumber \\
& & \int_0^\infty d\vf \vf \ff(\vf)\log{\left(1+\frac{p^2_{max}\vf^4}{G^2
M^2} \right)}.
\end{eqnarray}
\end{subequations}
The mean square velocity of the black hole in the multi-mass
case, for a Maxwellian distribution of field-star velocities,
is therefore predicted to be
\begin{equation}
\langle V^2\rangle_{mm} = \left({3C\over 2A}\right)_{mm} = 
3\left({\tilde m}\over{M}\right) \sigma^2,
\end{equation}
i.e. $\tilde m/m$ times its value in the case of a single
perturber mass $m$,
where the variable $\tilde m$ is
\begin{equation}
\tilde m \equiv {\int n(m) m^2 dm\over\int n(m) m dm}.
\label{eq:mtilde}
\end{equation}
This prediction will be tested below via $N$-body integrations.

Brownian motion results in a time-dependent displacement of the
black hole from its otherwise central location in a galaxy.
The $N$-body integrations presented here were judged to be too short
to extract useful information about the amplitude of this
displacement (as opposed to the velocity), 
and in what follows, we focus on changes in velocity.
We will return to the question of Brownian {\it displacements}
in a subsequent paper.

\section{Results: Single-Mass Models}

\begin{figure*}
\includegraphics[scale=0.75,angle=-90.]{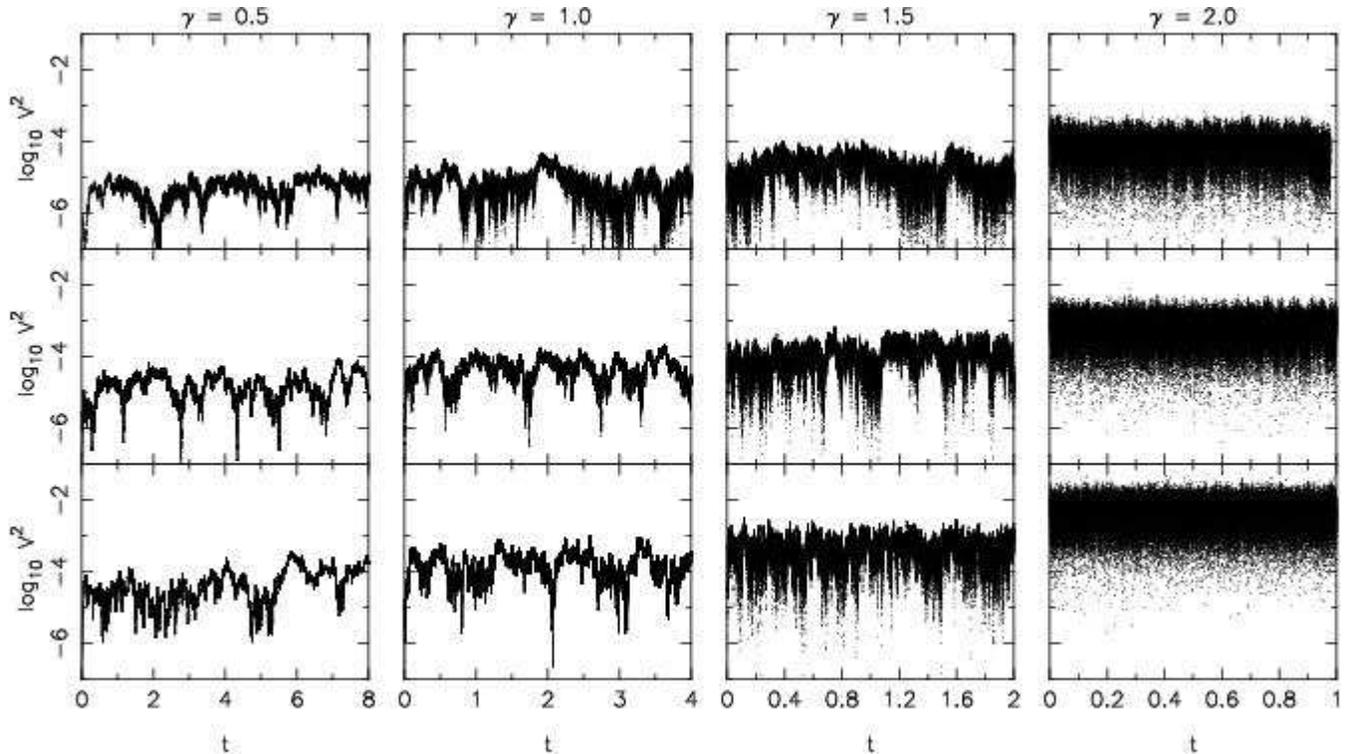}
\caption{
Time dependence of black hole velocity in a set of integrations.
From top to bottom, the black hole mass is 
$10^{-1}$, $10^{-2}$ and $10^{-3}$ in units of the galaxy
mass.
\label{fig:voft}
}
\end{figure*}

Figure~\ref{fig:voft} illustrates the time dependence
of the black hole's velocity in a representative set
of integrations.
The ``frequency'' of the motion increases with increasing
$\gamma$, as expected, since the period of small oscillations
about the potential center is smaller for larger $\gamma$.
However none of these plots exhibits the quasi-harmonic motion
characteristic of particles wandering in a constant-density
core (e.g. Miller \& Smith 1992).

\begin{figure}
%\epsscale{1.0}
\plotone{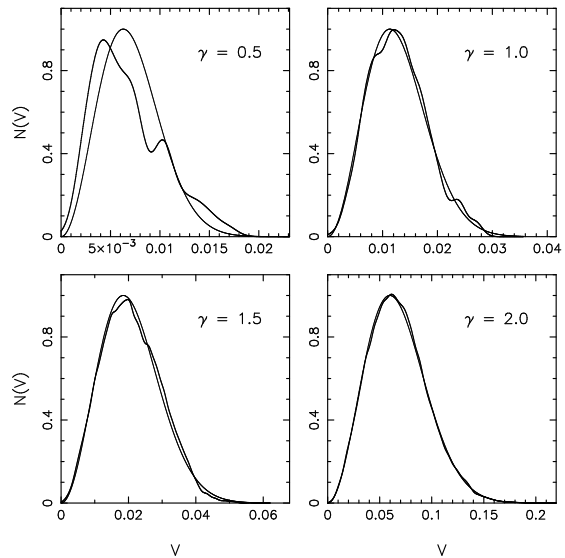}
\caption{
Measured distribution of black hole velocities in four integrations
with $M=10^{-3}$.
Thin curves show Maxwell-Boltzmann distributions, 
computed as described in the text.
\label{fig:MB}
}
\end{figure}

Brownian motion of a particle in a classical gas obeys
a Maxwell-Boltzmann distribution.
As shown above (\S4), the same is predicted by the 
Fokker-Planck equation for a massive particle moving in
response to gravitational perturbations in an infinite
homogeneous background of stars, as long
as the dynamical friction coefficient obeys
Hooke's law at low velocities.
However the equations derived above embody
many approximations that are violated in
the $N$-body models and in real galaxies.
Hence it is interesting to directly compute the
time-averaged velocity distribution of the
massive particle in the $N$-body integrations.
Figure~\ref{fig:MB} shows $N(V)$ for 
integrations with $M=10^{-3}$ and
four different values of $\gamma$.
The velocity of the black hole particle
was sampled at fixed intervals,
ranging from $\Delta t=0.002$ for $\gamma=0.5$
to $\Delta t = 0.00003$ for $\gamma=2.0$.

The empirical velocity distributions 
in Figure~\ref{fig:MB} are compared with the 
Maxwell-Boltzmann distribution,
\begin{equation}
N(V)dV = 4\pi V^2
\left(2\pi\langle V^2\rangle/3\right)^{-3/2}
\exp\left(-3V^2/2\langle V^2\rangle\right)dV.
\label{eq:MB}
\end{equation}
The quantity $\langle V^2\rangle$ in equation (\ref{eq:MB})
 was computed directly from the time
series of stored velocities; it was not adjusted to
increase the goodness of fit of equation (\ref{eq:MB}) 
to the measured $N(V)$.
Neverthless, the fit of the measured $N(V)$'s to the Maxwell-Boltzmann
distribution is clearly good,
and nearly perfect  in the case of $\gamma=2$;
in this model, the black hole particle experienced the
largest number of velocity updates.
We verified that the bumps and wiggles in the plots
of $N(V)$ for $\gamma=0.5$ and $\gamma=1.0$ were 
essentially random,
by constructing $N(V)$ separately for the first
and second halves of the integration intervals.

\begin{figure}
%\epsscale{1.0}
\plotone{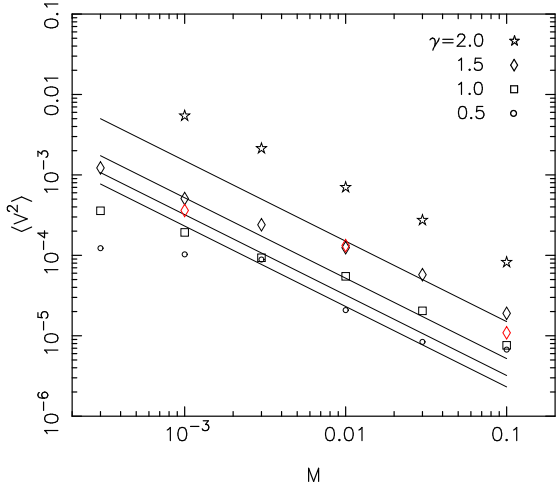}
\caption{
Mean square velocity of the black hole particle as given in
Table 2 for the various $N$-body integrations.
Lines show the ``equipartition'' relations, $\eta=1$.
Red diamonds are from Dorband et al. (2003).
\label{fig:vsq1}
}
\end{figure}

\begin{figure}
%\epsscale{1.0}
\plotone{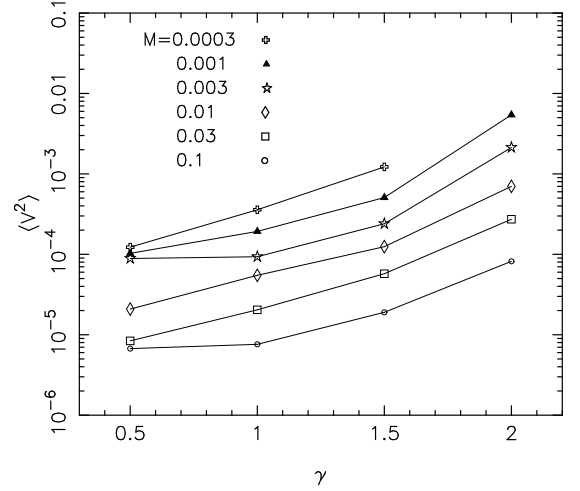}
\caption{
Mean square velocity of the black hole particle for the
$N$-body integrations listed in Table 2.
These are the same data as in Figure~\ref{fig:vsq1}, 
but plotted to show the dependence on $\gamma$.
\label{fig:vsq2}
}
\end{figure}

If we assume that the time-averaged $N(V)$ is well described 
by the Maxwell-Boltzmann distribution,
then the only quantity required to specify $N(V)$ is 
$\langle V^2\rangle$.
We expect $\langle V^2\rangle$ to be close to the 
``energy equipartition'' value at which the time-averaged
kinetic energy of the black hole equals the mean kinetic
energy of nearby stars.
We write this as
\begin{equation}
\langle V^2\rangle = 3\eta {m\over M}\sigma_p^2
\label{eq:equipart}
\end{equation}
where $\sigma_p$ is the peak
value of the 1D stellar velocity dispersion, as defined
in \S 2, and $\eta$ is a parameter whose value is 
expected to be close to one.
Values of $\langle V^2\rangle$ 
are given in Table~\ref{tab:two}, and plotted in
Figures~\ref{fig:vsq1} and~\ref{fig:vsq2} as functions
of $M$ and $\gamma$.
These values were computed by sampling the stored
velocities at small fixed intervals.
The lines in Figure~\ref{fig:vsq1} show the ``equipartition''
relations, $\eta=1$.

Two results are apparent from Figures~\ref{fig:vsq1} 
and~\ref{fig:vsq2}.

1. There is a clear dependence of $\langle V^2\rangle$
on $M$, although less steep than the expected
$\langle V^2\rangle \propto M^{-1}$ dependence.

2. For a fixed black hole mass, 
the fractional discrepancy between the measured 
$\langle V^2\rangle$ and the 
``equipartition'' value 
is an increasing function of $\gamma$.

\begin{figure}
\epsscale{1.0}
\plotone{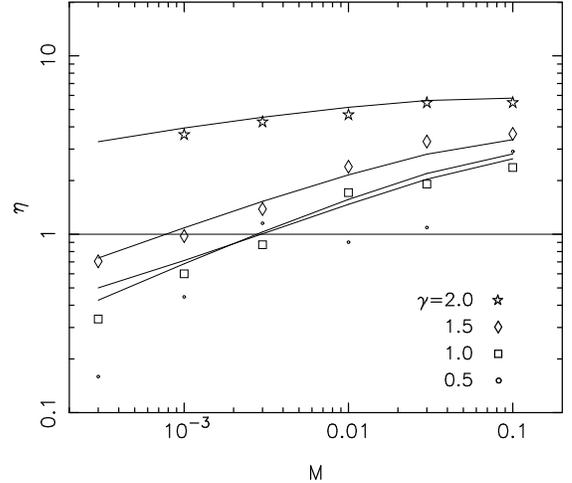}
\caption{
The quantity $\eta$ (equation~15) that measures departures from 
``equipartition'' in the $N$-body models.
Lines are the values of $\eta$ predicted if the black hole is
assumed to be in energy equipartition with stars lying 
at distances $\le 0.6r_h$ from it.
\label{fig:eta}
}
\end{figure}

Another way to present these results is to 
interpret equation (\ref{eq:equipart}) as defining
the parameter $\eta$, which then measures departures from
``equipartition'' as a function of $M$ and $\gamma$.
Figure~\ref{fig:eta} plots $\eta$,
so defined, for these integrations; the $\eta$ values 
are also given in Table 2.
As the two previous figures indicated, 
$\eta$ increases systematically with $M$ and $\gamma$,
reaching values as large as $\sim 5.5$ for $(M,\gamma)=(0.1,2.0)$.
For small ($M,\gamma$), $\eta$ drops  below one, to 
values as low as $\sim 0.2$.

\begin{deluxetable}{lll}
\tablewidth{0pt}
\tablecaption{$\log_{10}\eta=A+B\log_{10} M$\label{tab:coef}}
\tablehead{
\colhead{$\gamma$} & \colhead{$A$} & \colhead{$B$} 
}
\startdata
 0.5& 0.844 & 0.420 \\
 1.0& 0.801 & 0.345 \\
 1.5& 0.934 & 0.307 \\
 2.0& 0.853 & 0.093 \\
\enddata
\end{deluxetable}

The  dependence of  $\log\eta$ on $\log M$ is nearly
linear for fixed $\gamma$ (Figure~\ref{fig:eta}).
We carried out least-squares fits to  find the coefficients
of $\log_{10}\eta \approx A + B\log_{10} M$.
The results of  the fits are given in Table~\ref{tab:coef}.
These results imply that -- in a galaxy with otherwise
fixed properties -- the mean square velocity of the black
hole  scales as $\sim M^{-0.6}$ in low-density nuclei
and $\sim M^{-0.9}$ in  high-density nuclei.
Ignoring the effect of the black hole on the stellar motions
would lead to the prediction $\langle V^2\rangle\propto M^{-1}$.

There is a natural way to understand these results.
The steady-state velocities of the stars in these galaxy models
are influenced by the presence of  the black hole 
(Figure~\ref{fig:dehnen}), very strongly at distances $\lap r_h$.
The Brownian motion of the black hole depends in turn on
perturbations from these fast-moving stars.
The more massive the black hole, the hotter the  surrounding
stellar fluid, and the greater the expected deviation
of the black hole's Brownian velocity from the value that  would  have been
predicted based on the properties of a model with the same
density structure but lacking a black hole.
These deviations should also be an increasing function of
$\gamma$, since for larger $\gamma$, the number of high-velocity
stars near the black hole  is larger.

Suppose we assume that the black hole's kinetic
energy is in equipartition with
the stars in some region around it.
Define $\tilde{\sigma}^2$ to be the 1D, mean square stellar velocity 
within this  region, which has radius $\tilde r$.
The black hole's Brownian velocity would then be
\beq
\langle V^2\rangle = 3{m\over M} \tilde{\sigma}^2,
\label{eq:tilde}
\eeq
It is understood that $\tilde{\sigma}^2$ in equation (\ref{eq:tilde})
includes the
effect of the black hole on the equilibrium stellar motions.
We might guess, based on the $N$-body results,
that the best choice for $\tilde r$ is
some multiple of the black hole's radius  of influence $r_h$,
$\tilde r = F\times r_h$.

Before testing this hypothesis, we make two remarks.

\noindent 1. The standard definition of $r_h$ is $GM/\sigma^2$.
This definition is adequate for galaxies where $\sigma(r)$
is approximately constant near the center.
For the Dehnen models, this is only true for $\gamma=2$,
and it is never true in models containing a central black hole.
In \S3, we generalized the definition of $r_h$ to be the
radius containing a mass in stars equal to twice the black
hole mass.
Thus, at $r=r_h$, the gravitational force from the  
black hole is one-half that due to the stars.
We retain that definition in what follows.

\noindent 2. Above we defined $\tilde\sigma^2$ as the 
mean value of $\sigma^2$ within some radius, 
including the effect of the black
hole's gravity on the stellar motions.
However for $\gamma\ge 2$, this quantity {\it diverges}
as a function of  the lower integration limit, 
due to the  $\sigma^2\propto r^{-1}$ increase in velocities
near the black hole.
Nevertheless the divergence is only
logarithmic as a function of the lower integration limit
 when $\gamma=2$.
In our $N$-body models, a natural choice for  the lower
integration limit is $\epsilon$, the softening length.
We made this choice in what follows.

Figure~\ref{fig:eta} shows the predictions of equation (\ref{eq:tilde})
compared with the $N$-body data, setting $F=0.6$,
i.e. when $\tilde{\sigma}^2$ is computed as an average within
$0.6 r_h$.
In this plot, the ordinate is the value of $\eta$
that would have been measured if the black hole's Brownian velocity 
were given by equation (\ref{eq:tilde}),
i.e. $\eta=\tilde\sigma^2/\sigma_p^2$.
The agreement with the data is quite good,
considering the simplicity of the prescription, 
the wide range in density structure of the models,
and the fairly arbitrary definition of $r_h$.
The worst fit is for $\gamma=0.5$, however these
points exhibit a large scatter, probably reflecting
that the $N$-body integration time was barely long
enough to provide a robust sampling of the massive
particle's motion (Fig.~\ref{fig:voft}).

Other definitions  of $\tilde r$ were tried 
(e.g., a fixed value in model units)
but none was found that reproduced the measured values of 
$\eta$ so well.

We carried out least-squares fits against the data of
Figure~\ref{fig:eta} to determine the best value of $F$ for each $\gamma$.
For $\gamma=(0.5,1,1.5,2)$, we found best-fit $F$'s of
$(0.77,0.76,0.52,0.66)$.
The fit to the $\gamma=2$ data was nearly perfect when the
optimal $F$ was used.
We conclude that the ``black holes'' in our  simulations
are in approximate energy equipartition with the stars 
that lie in a
sphere of radius  $(0.67\pm 0.10) r_h$ around them.
Excluding the rather poor fit for $\gamma=0.5$, this
becomes $(0.65\pm 0.10) r_h$.

We can develop a simple model that reproduces these results.
We suppose that the black hole's motion can be broken into two
pieces:
\beq
\langle V^2\rangle = \langle V_1^2\rangle + \langle V_2^2\rangle.
\eeq
The first piece, $\langle V_1^2\rangle$, 
refers to the motion of the center of mass of the system
consisting of the black hole and the stars bound to it.
The second piece, $\langle V_2^2\rangle$, is the mean
square velocity of the black hole with respect to 
the center of mass of the bound system.

Our model requires that we first identify which stars
are ``bound'' to the black hole.
Clearly, stars with apocenter distances $r_+\ll r_h$ are bound
to the black hole, and
stars with $r_+\gg r_h$ are
not.
However stars with $r_+\approx r_h$ can neither be said
to be bound or unbound.
We will call a star ``bound'' if
\beq
r_+ < F' r_h.
\eeq
Here $F'\approx 1$ is a free parameter similar to the factor
$F$ defined above.
According to this definition, the gravitational force on the
``bound'' stars comes predominantly from the black hole.

Let the mass in bound stars be $M_b$.
Then the effective mass of the bound system 
(black hole plus stars) is
\beq
M_{eff} = M + M_b.
\eeq
The remaining (unbound) stars define a core with
some characteristic density and velocity dispersion.
Define the latter to be $\sigma_u$.
We compute $\sigma_u$ by taking an average over the
unbound stars within a region of radius $2F'r_h$.

Given these definitions, the contribution to the
black hole's motion from the unbound stars is 
given by Chandrasekhar's theory (\S 4):
\beq
\langle V_1^2\rangle \approx 3{m\over M_{eff}}\sigma_u^2.
\eeq
In other words, we expect the bound system, of mass $M_{eff}$,
to act like a single particle in energy equipartition with the 
unbound stars in the core.

The motion of the black hole with respect to the center of
mass of the bound system can be computed by setting
\beq
M{\bf V}_2 + m \sum_{\rm bound} {\bf v}_i = 0,
\eeq
where ${\bf V}_2$ and ${\bf v}_i$ are the velocity of the black hole
and of a bound star with respect to the center of mass of the 
bound system.
Ignoring correlations between the motions of the
bound stars, we find
\beq
\langle V_2^2\rangle = {3M_bm\over M^2}\sigma_b^2
\eeq
with $\sigma_b$ the velocity dispersion of the bound stars.

Combining the two pieces gives
\beq
\langle V^2\rangle \approx {3m\over M}\left({M\over M+M_b}\sigma_u^2
+ {M_b\over M}\sigma_b^2\right).
\eeq
Since $M_b\approx M$, this is roughly
\beq
\langle V^2\rangle \approx {3m\over M}\left({\sigma_u^2\over 2}
+\sigma_b^2\right).
\label{eq:app2}
\eeq
For $F'\approx 1$, $\sigma_b$ is of the same order or greater
than $\sigma_u$ for most $(\gamma,M)$, hence the quantity in parentheses on the
right hand side of equation (\ref{eq:app2}) is comparable
to $\tilde{\sigma}^2$.
Thus our model naturally reproduces the dependence of
$\langle V^2\rangle$ observed in the $N$-body simulations.

\begin{figure}
\epsscale{1.0}
\plotone{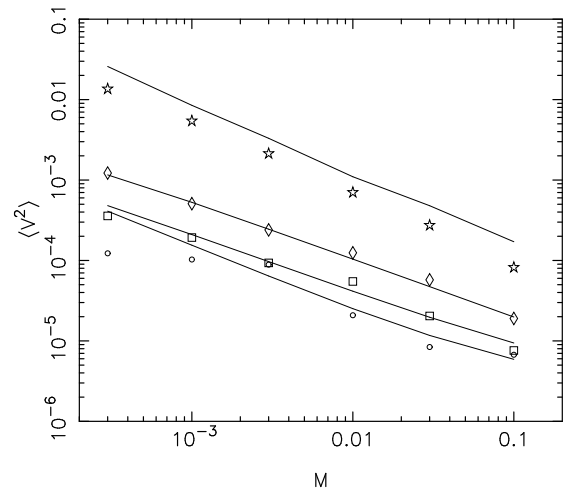}
\figcaption{
Predicted values of $\langle V^2\rangle$ 
based on the model of \S 5.
Symbols show the $N$-body data, as in Figure 4.
\label{fig:simple}
}
\end{figure}

Figure~\ref{fig:simple} shows predicted values of $\langle V^2\rangle$
for $F'=1$.
The match with the data in Figure~\ref{fig:vsq1} is quite good.

\section{Results: Multi-Mass Models}

Stars in real galaxies do not all have the same mass. 
We carried out an additional set of $N$-body integrations based on the
same galaxy models as before, but generating the mass of
each ``star'' particle from a \cite{Salpeter:55} mass function,
\begin{equation}
n(m)dm \varpropto m^{-(1+\alpha)}dm,\ \ \ \ m_1 \leq m \leq m_2,
\label{eq:salp}
\end{equation}
with $\alpha=1.35$.
We chose $m_2/m_1=100$; the values of $m_1$ and $m_2$ are then
fixed by the requirement that the mean stellar mass
equal $N^{-1}$.
Each of these integrations used $M=10^{-2}$ and $N=10^6$.
Because these galaxy models contain a small number of massive
stars (those at the high-mass tail of the distribution),
we expect the statistical variance of the results to
be greater than in the single-mass runs.
Hence we carried out $10$ integrations of each model 
using different
seeds to initialize the random number generator for the stellar masses.
The $\langle V^2\rangle$ values given in Table~\ref{tab:multi}
are averages over these 10 runs.
We also used shorter integration times than in the single-mass
integrations in order to reduce the effects of mass segregation: 
$T=(2,1,0.5,0.25)$ for
$\gamma=(0.5,1,1.5,2)$ respectively.
We give in Table~\ref{tab:multi} $\sigma_{\langle V^2\rangle}$,
the variance in the measured $\langle V^2\rangle$ values for
the 10 runs, and ratio between the mean $\langle V^2\rangle$ and the value
measured in the single-mass runs for the same ($\gamma,M,N$).
We computed the latter quantities over the same time intervals
used for the multi-mass integrations.

\begin{figure}
\epsscale{1.0}
\plotone{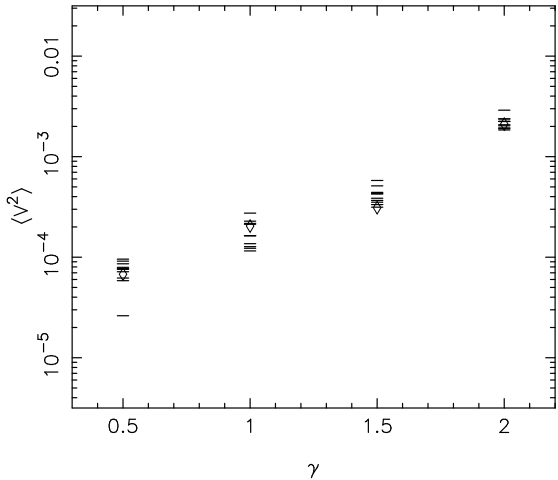}
\figcaption{
Mean square black hole  velocity in the multi-mass integrations.
For each value of $M$, ten $N$-body integrations were carried
out using different seeds for the random number generator
for the stellar masses; the results for each
integration are indicated by $-$.
The $\diamond$ symbols show the predicted values based
on the theory in \S 4.
\label{fig:multi}
}
\end{figure}

As shown in  \S 4, Chandrasekhar's theory allows us to 
predict a relation between
the black hole's mean square velocity in the multi-mass 
and single-mass cases.
The former is predicted to  be larger than the latter
by a factor $\tilde m/m$, where $\tilde m$ is defined
in equation (\ref{eq:mtilde}).
Substituting equation (\ref{eq:salp}) into equation
(\ref{eq:mtilde}) gives
\begin{equation}
\tilde m = \left({1-\alpha\over 2-\alpha}\right) m_1
\left[{\left({m_2\over m_1}\right)^{2-\alpha}-1\over
\left({m_2\over m_1}\right)^{1-\alpha}-1}\right].
\end{equation}
Setting $\alpha=1.35$ and $m_2/m_1=100$ yields
$\tilde m = 12.7 m_1  = 4.10\overline{m}$ where $\overline{m}=N^{-1}$
is the mean stellar mass.
Hence, in the multi-mass integrations, Chandrasekhar's theory predicts 
a mean square velocity for the black hole that is roughly four times 
larger than in the single-mass integrations.

The final column of Table~\ref{tab:multi} gives the measured
values of this ratio, and 
Figure~\ref{fig:multi} 
shows the
10 measured values of $\langle V^2\rangle$ for each value
of $\gamma$, compared with $4.1$ times the measured value in the
corresponding single-mass run.
While the variation  from run to run is large,
the mean increase in $\langle V^2\rangle$ is 
reasonably close to the predicted factor.

\begin{deluxetable}{llll}
\tablewidth{0pt}
\tablecaption{Multi-mass $N$-body integrations\label{tab:multi}}
\tablehead{
\colhead{$\gamma$} & \colhead{$\langle V^2\rangle$} & 
\colhead{$\sigma_{\langle V^2\rangle}$} & \colhead{ratio}
}
\startdata
0.5 & $ 6.58\times 10^{-5}$ & $2.8\times 10^{-5}$ & 3.98\\
1.0 & $ 1.60\times 10^{-4}$ & $7.0\times 10^{-5}$ & 3.22\\
1.5 & $ 3.77\times 10^{-4}$ & $1.4\times 10^{-4}$ & 4.98\\
2.0 & $ 1.99\times 10^{-3}$ & $6.9\times 10^{-4}$ & 3.87\\
\enddata
\end{deluxetable}

\section{Discussion}

Our $N$-body integrations show that a black hole
at the center of a dense stellar system responds to
perturbations  from passing stars by achieving  
a time-averaged kinetic energy given by
\beq
{1\over 2}M\langle V^2\rangle \approx {3\over 2}\tilde{m}\tilde{\sigma}^2
\label{eq:result}
\eeq
where $\tilde{\sigma}^2$ is the 1D, mean square 
stellar velocity within a region $r\lap 0.6 r_h$
around the black hole,
$r_h$ is the black hole's influence radius
(defined as the radius containing a mass in stars equal
to twice the black hole's mass),
and $\tilde m$ is an effective stellar mass
defined via equation (\ref{eq:mtilde});
in the case of a delta-function mass spectrum, 
$\tilde m$ is equal to the stellar mass.
Equation (\ref{eq:result}) was found to provide a reasonably
accurate description of the $N$-body integrations in galaxy
models with power-law nuclei, $\rho\propto r^{-\gamma}$,
$0.5\le\gamma\le 2$, and
for black hole masses in the range $10^{-4}\le M/M_{gal}\le 10^{-1}$.
A more accurate description of the $N$-body results is given
by equation (\ref{eq:equipart}) (with $m$ replaced by $\tilde {m}$)
and Table 3,
or simply by the measured values in Tables 2.
The detailed distribution of black hole velocities,
$N(V)$, was found to be essentially indistinguishable
from a Maxwell-Boltzmann distribution (Figure~\ref{fig:MB}).

As discussed above (\S 5),
equation (\ref{eq:result}) has a simple physical interpretation:
the presence of the black hole deepens the galaxy's central potential
and increases the velocities of stars in its vicinity,
and these high-velocity stars provide in turn the dominant perturbations
that determine the amplitude of the black hole's Brownian velocity.
The dependence of $\langle V^2\rangle$ on $M$ is therefore
less steep than $M^{-1}$, since increasing $M$ also increases
the velocities of nearby stars.

If we accept equation (\ref{eq:result}),
it is straightforward to derive simple analytic expressions for
the dependence of $\langle V^2\rangle$ on $M/M_{gal}$ and $\gamma$.
For the Dehnen models studied here, the black hole's  influence
radius is given by
\beq
r_h = {(2M)^{1/(3-\gamma)}\over 1-(2M)^{1/(3-\gamma)}}
\approx (2M)^{1\over(3-\gamma)}, \ \ \ \ M\ll 1
\eeq
in model units ($G=M_{gal}=a=1$).
The quantity $\tilde{\sigma}^2$ that appears in equation (\ref{eq:result})
is defined as a number-weighted mean
within a distance $F\times r_h$ from the black hole, where $F\approx 0.6$
(\S 5).
Here we can make use of the analytic expressions for
$\sigma^2(r)$ given by \cite{Dehnen:93} and \cite{Tremaine:94}.
We write
\beq
\sigma^2(r) = \sigma_1^2(r) + M\sigma_2^2(r)
\eeq
where $\sigma_1^2$ is the contribution to $\sigma^2$ from the
stellar potential, and $\sigma_2^2$ is the additional component
due to the presence of the black hole.
For  $1<\gamma\le 2$ and at radii $\lap r_h$, we have
\beq
\sigma_1^2(r) \approx {r^{2-\gamma}\over 2(\gamma-1)},\ \ \ \ 
\sigma_2^2(r) = {M\over (1+\gamma)}{1\over r}.
\label{eq:app}
\eeq
Averaging over a sphere of radius $Fr_h$,
and expressing  the result in physical units,
we find
\begin{subequations}
\begin{eqnarray}
\tilde{\sigma}^2 &\approx& H(\gamma,F)
\left({GM_{gal}\over a}\right) 
\left({M\over M_{gal}}\right)^{(\gamma-2)/(\gamma-3)}, \\
H(\gamma,F) &=& 2^{1/(\gamma-3)}(3-\gamma)F^{-1} \nonumber \\
& & \left[{(1+\gamma)(2-\gamma)F^{3-\gamma} + (\gamma-1)(5-2\gamma)\over (\gamma^2-1)(5-2\gamma)(2-\gamma)}\right].
\end{eqnarray}
\end{subequations}
(Note the divergence when $\gamma=2$, discussed in \S 5.)
Setting $F=0.6$ (\S 5) gives $H\approx 2.3\pm 0.4$ for $\gamma$ in the range
 $1.2\le\gamma\le 1.8$.
The black hole's mean square velocity is then predicted to be,
via equation (\ref{eq:result}),
\beq
\langle V^2\rangle \approx 7\left({GM_{gal}\over a}\right) \left({\tilde{m}\over M_{gal}}\right) \left({M_{gal}\over M}\right)^{1/(3-\gamma)},\ \ \ \ 1\lap\gamma\lap 2.
\label{eq:approx}
\eeq
When translated back into model units, this is
$\langle V^2\rangle\approx 7N^{-1}M^{-1/(3-\gamma)}$,
which is a tolerable fit to the $\langle V^2\rangle$ values plotted
in Figures~\ref{fig:vsq1} and \ref{fig:simple}.
The analytic expressions for $\langle V^2\rangle$ when
$\gamma$ is in the range $0.5\le\gamma\le 1$
are more complicated but still imply 
$\langle V^2\rangle \propto M^{-1/(3-\gamma)}$ for small $M/M_{gal}$.

The fact that $\langle V^2\rangle$ falls off less steeply
than $M^{-1}$ with increasing mass is perhaps the first 
clear indication that gravitational 
Brownian motion differs in a significant way from its fluid analog,
for which $\langle V^2\rangle\propto M^{-1}$.
It would be of interest to refine the $M$-dependence.
Extending the $N$-body integrations to longer times would
reduce the noise but increase the possibility of systematic
errors due to collisional evolution of the models.
An alternative would be to carry out a large number of shorter
integrations and average the results, as was done here
in the multi-mass studies (\S 6).

When making predictions about the expected amplitude
of the Brownian velocity of black holes in real galaxies,
an expression like equation (\ref{eq:approx}) is
not ideal since it contains the terms $M_{gal}$ and $a$
that depend on the large-radius properties of the galaxy.
A less model-dependent way to present these results is in terms
of the central properties of the galaxy.
Here we return to equation (\ref{eq:equipart}) and to
the empirical relation established between $\eta$ and $M$:
\begin{subequations}
\begin{eqnarray}
\langle V^2\rangle &=& 3\eta {\tilde{m}\over M}\sigma_p^2,\\
\log_{10}\eta&\approx& A + B\log_{10} M
\end{eqnarray}
\label{eq:final}
\end{subequations}
where the best-fit coefficients $A$ and $B$ as derived
from the $N$-body integrations are given in Table 3.
Equations (\ref{eq:final}) give $\langle V^2\rangle$ in terms
of $\sigma_p$, the peak value of $\sigma$ measured outside of the
black hole's influence radius (Figure~\ref{fig:dehnen}, Table 1).
This quantity is easily accessible via ground-based observations
for many galaxies.

As an example, we consider the nucleus of the Milky Way.
The peak velocity dispersion is
$\sigma_p\approx 150$ km s$^{-1}$ \citep{Kent:92}.
For the stellar density profile at the center of the Milky
Way bulge, \cite{Genzel:03} find $\rho\propto r^{-\gamma}$
with $\gamma\approx 1.4\pm 0.1$, consistent with the value 
$\gamma=1.5$ used here in some of the $N$-body integrations.
Finally, the black hole mass is $3.4\pm 0.5\times 10^6\msun$
\citep{Schoedel:03}.
Since the Milky Way black hole falls on the tight scaling
relations defined by the other secure black hole masses
\citep{FM:00,MH:03}, we assume that 
$M/M_{gal}\approx 1.25 \times 10^{-3}$,
the mean ratio of black hole mass to bulge mass
defined by these galaxies \citep{MF:01}.
From Table 4 or Figure~\ref{fig:eta}, 
we then have $\eta\approx 1.5$ and
\begin{eqnarray}
V_{rms} &\approx& 0.17\ {\rm km\ s}^{-1} 
\left({\tilde{m}\over 1\msun}\right)^{1/2} 
\left({M\over 3.4\times 10^6\msun}\right)^{-1/2} \nonumber \\
& & \times \left({\sigma_p\over 150\ {\rm km\ s}^{-1}}\right).
\end{eqnarray}
The characteristic stellar mass $\tilde{m}$ 
(equation \ref{eq:mtilde})
depends on the poorly-known mass function for stars in the
Galactic nucleus but $\tilde{m}$ is probably
of order $\msun$ \citep{Genzel:03}.
Hence we predict $V_{rms}\approx 0.2$ km s$^{-1}$.

Current limits on the motion of the Milky Way black hole
are based on proper motion measurements of images of Sagittarius A$^*$
($V_{pm}\lap 8$ km s$^{-1}$; \cite{Reid:03}).
Such measurements require observations over an extended
period of time; in the case of the Reid et al. (2003)
study, the baseline was $\sim 7$ yr.
The measured velocity is derived from the integrated
displacement.
If the black hole's velocity changes significantly
during this time, the distribution of measured
velocities will be different than the distribution
of instantaneous velocities.
To evaluate the importance of this effect for
proper motion measurements in the Galactic center,
we used our $N$-body results to compute displacements
of the black hole particle over finite times, then
computed distributions of the proper motion velocities
that would be measured from these displacements.
(We note that the time scale over which the massive
particle's velocity changes is predicted to be
{\it independent} of the number of perturbers for
a given mean perturber density, based on the equations
in \S 4.)
After scaling to the Galactic center, we found almost
no dependence of the measured (proper motion) velocities
on the measurement
baseline, for times up to $\sim 10^3$ yr. 
We therefore base the following discussion on the
instantaneous velocity distribution $N(V)$.

Given that $N(V)$ is expected to be a Maxwell-Boltzmann
distribution (equation \ref{eq:MB}), we can 
use the estimate of $V_{rms}$ to compute
the probability that a random velocity measurement
will exceed any value $V_0$.
We first convert the velocity distribution of equation
(\ref{eq:MB}) into a distribution over 2D velocities
$V_{pm}$ in the plane of the sky.
Transforming, we find
\beq
N(V_{pm})dV_{pm} = (\langle V^2\rangle/3)^{-1} 
V_{pm}e^{-3V_{pm}^2/2\langle V^2\rangle} dV_{pm}.
\label{eq:prob1}
\eeq 
The probability of measuring $V_{pm}$ to be
greater than $V_0$, given $\langle V^2\rangle$,
is then just $\exp(-3V_0^2/2\langle V^2\rangle)$.
If $\langle V^2\rangle = (0.17\ {\rm km\ s}^{-1})^2$,
the expected value for the Milky Way black hole,
the probability that $V_{pm}$ will exceed 
$V_0=(0.1,0.2,0.3,0.4,0.5)$ km s$^{-1}$
is $(0.60, 0.13,9.4\times 10^{-3}, 2.5\times 10^{-4},
2.3\times 10^{-6})$.
In order to have a reasonable chance of detecting the black
hole's motion, 
a velocity resolution better than $0.3$ km s$^{-1}$
will be required.

Once a bona fide measurement of $V_{pm}$ has been made, 
the black hole's velocity 
can be converted into an estimate of its mass.
Here we make use of Bayes's theorem:
\beq
P(M|V_{pm}) = {P(M)P(V_{pm}|M)\over \int dM'P(M')P(V_{pm}|M')}.
\eeq
where $P(a|b)da$ denotes the probability, given $b$, of 
measuring $a$ in the range $a$ to $a+da$, and 
$P(M)dM$ is the prior probability that $M$ lies in the 
range $M$ to $M+dM$.
Here ``prior'' means ``given all knowledge prior to the
measurement of $V_{pm}$.''
A completely uninformed prior would be $P(M)=$ constant;
another standard choice is $P(M)\propto M^{-1}$;
and of course one could use information about $M$
from other sources (e.g. \cite{Schoedel:03}) to construct
even more informed priors.

Adopting the maximally-uninformed prior, 
and using equations 
(\ref{eq:final})
and (\ref{eq:prob1}) 
to write $P(V_{pm}|M)$, we find the probability
distribution of $M$ given a measured $V_{pm}=V_0$ to be
\beq
P(M|V_0)dM = {V_0^4\over 4\tilde{m}^2\eta^2\sigma_p^4} 
Me^{-MV_0^2/2\tilde{m}\eta\sigma_p^2} dM
\label{eq:p1}
\eeq
and the probability that $M\ge M_0$ is
\beq
P(M\ge M_0|V_0) = \left(1+{M_0V_0^2\over 2\tilde{m}\eta\sigma_p^2}\right)e^{-M_0V_0^2/2\tilde{m}\eta\sigma_p^2}.
\label{eq:upper}
\eeq
(In deriving these equations we have ignored the relatively weak dependence
of $\eta$ on $M$, Table 4.
Including that dependence is straightforward but results in non-analytic
expressions for $P$.)
We define the ``best estimate'' ${\cal M}$ of the black hole's
mass to be the median value from this distribution.
This is
\beq
{\cal M} = {3.36\tilde{m}\eta\sigma_p^2\over V_0^2}.
\eeq
We propose to call this the ``Brownian mass estimator.''
For the Milky Way black hole ($\sigma_p\approx 150$ km s$^{-1}$,
$\eta\approx 1.5$), 
the estimated mass is
\beq
{\cal M} \approx 2.8\times 10^6\msun \left({\tilde{m}\over 1\msun}\right)
\left({V_0\over 0.2\ {\rm km\ s}^{-1}}\right)^{-2}.
\label{eq:MW}
\eeq
The $90\%$ confidence intervals on the mass are given by the
values of $M_0$ for which $P(M\ge M_0|V_0)=(0.05,0.95)$, or
\beq
0.211 {\cal M} \le M \le 2.82 {\cal M}.
\eeq

Unlike mass estimators based on the statistics of a large
sample,
the uncertainty associated with Brownian
mass estimators is irreducible, and it probably
makes more sense to interpret a measured $V_{pm}$
as defining a probability distribution for $M$ via equation
(\ref{eq:upper}) than a most-likely value via equation (\ref{eq:MW}).
For example, if $V_{pm}$ for the Milky Way black hole
were measured to be $0.1$ km s$^{-1}$, 
the probability that its mass exceeds 
$(10^6, 10^7, 10^8)\msun$ would be
$(99.0\%,56.4\%,0.00058\%)$.

\begin{figure}
\epsscale{1.0}
\plotone{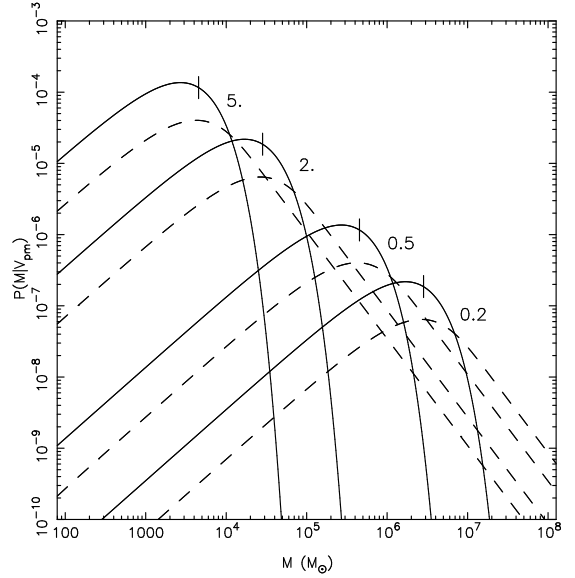}
\figcaption{
Probability distribution for the mass of the Milky Way black hole,
based on measurement of a proper motion velocity (solid curves),
or based on determination of an upper limit to the proper-motion
velocity (dashed curves),
for four values of $V_{pm}$ in km s$^{-1}$.
Tick marks indicate median values.
\label{fig:prob}
}
\end{figure}

Figure~\ref{fig:prob} plots $P(M|V_0)$ in the case of the Milky Way black hole
($\sigma_p=150$ km s$^{-1}$, $\eta=1.5$)
for four values of $V_0$.
Also shown (by tick marks) are the median values ${\cal M}$.

If we adopt instead the ``informed'' prior,
$P(M)\propto M^{-1}$, then
\begin{subequations}
\begin{eqnarray}
P(M|V_0)dM &=& {V_0^2\over 2\tilde{m}\eta\sigma_p^2} 
e^{-MV_0^2/2\tilde{m}\eta\sigma_p^2}, \\
P(M\ge M_0|V_0) &=& e^{-M_0V_0^2/2\tilde{m}\eta\sigma_p^2}
\end{eqnarray}
\label{eq:p2}
\end{subequations}
and the Brownian mass estimator becomes
\beq
{\cal M} = {1.39\tilde{m}\eta\sigma_p^2\over V_0^2}
\approx 1.2\times 10^6\msun \left({\tilde{m}\over 1\msun}\right)
\left({V_0\over 0.2\ {\rm km\ s}^{-1}}\right)^{-2}.
\eeq
The median mass is now smaller due to the prior which disfavors
high masses.

Finally, we ask what can be learned about the black hole
mass if only an upper limit on its velocity, $V_{pm}\le V_{up}$,  is available.
We consider the determination of an upper limit on $V_{pm}$
to be equivalent to the statement that any velocity in the interval
$0\le V_{pm}\le V_{up}$ is equally likely, and that velocities
greater than $V_{up}$ have zero probability.
The probability distribution for the mass is then
\beq
P(M|V\le V_{up}) = {\int_0^{V_{up}} P(M|V_0)dV_0\over\int_0^{V_{up}} dV_0}.
\eeq
Again adopting the uninformed prior for $P(M)$, we find
\begin{eqnarray}
P(M|V\le V_{up}) &=& {1\over 8M} {{\cal V}\over V_{up}} 
\Bigg\{3\sqrt{\pi}{\rm erf}\left({V_{up}\over{\cal V}}\right) - 2e^{-V_{up}^2/{\cal V}^2} \nonumber \\
& & \times \left[{V_{up}\over {\cal V}}\left(3+2{V_{up}^2\over{\cal V}^2}\right)\right]\Bigg\}
\end{eqnarray}
with
\beq
{\cal V}^2(M) = {2\tilde{m}\eta\sigma_p^2\over M}. 
\eeq
This function is plotted in Figure~ref{fig:prob} for various $V_{up}$,
using the same values of $\tilde{m}$ and $\sigma_p$
given above for the Milky Way.
The low-$M$ tail of the distribution is similar to that
of $P(M|V_0)$, but there is a more extended tail at high $M$
corresponding to the fact that low values of $V_{pm}$ are
(by assumption) equally as likely as $V_{up}$.

\bigskip

We thank the anonymous referee for comments that substantially
improved this paper.
This work was supported by grants
AST-0206031, AST-0420920 and AST-0437519 from the NSF, 
grant NNG04GJ48G from NASA,
grant HST-AR-09519.01-A from STScI,
grant I/80 041 GRACE from the Volkswagen Foundation,
by SFB439 of Deutsche Forschungsgemeinschaft,
and by INTAS grant IA-03-59-11.
We thank the Center for the Advancement of the Study of 
Cyberinfrastructure at RIT for their support.


\begin{thebibliography}{}


\bibitem[Backer \& Sramek(1999)]{Backer:99}
  Backer, D. C. \& Sramek, R. A. 1999,
  ApJ, 524, 805

\bibitem[Bahcall \& Wolf(1976)]{BW:76}
  Bahcall, J. N. \& Wolf, R. A. 1976,
  ApJ, 209, 214

\bibitem[Berczik et al.(2005)]{BMS:05}
  Berczik, P., Merritt, D. \& Spurzem, R. 2005,
  ApJ, 633, 680

\bibitem[Chatterjee, Hernquist, \& Loeb(2002a)]{CHL:02a}
  Chatterjee, P., Hernquist, L. \& Loeb,  A. 2002a,
  ApJ, 572, 371

\bibitem[Chatterjee, Hernquist, \& Loeb(2002b)]{CHL:02b}
  Chatterjee, P., Hernquist, L. \& Loeb,  A. 2002b,
  PRL 88, 121103

\bibitem[Chatterjee, Hernquist, \& Loeb(2003)]{CHL:03}
  Chatterjee, P., Hernquist, L. \& Loeb,  A. 2003,
  ApJ, 592, 32

\bibitem[Dehnen(1993)]{Dehnen:93}
  Dehnen, W. 1993,
  \mnras, 265, 250

\bibitem[Dorband, Hemsendorf, \& Merritt(2003)]{DHM:03}
  Dorband, E. N., Hemsendorf, M., \& Merritt, D. 2003,
  J. Comp. Phys., 185, 484

\bibitem[Ferrarese \& Merritt(2000)]{FM:00}
  Ferrarese, L. \& Merritt, D. 2000,
  \apj, 539, L9

\bibitem[Fukushige et al.(2005)]{FMK:05}
  Fukushige, T., Makino, J. \& Kawai, A. 2005,
  PASJ, 57, 1009


\bibitem[Genzel et al.(2003)]{Genzel:03}
  Genzel, R. et al. 2003,
  \apj, 594, 812

\bibitem[Kent(1992)]{Kent:92}
  Kent, S. M. 1992,
  \apj, 387, 181

\bibitem[Laun(2004)]{Laun:04}
  Laun, F. 2004,
  Masters Thesis, 
  Rutgers University

\bibitem[Lightman \& Shapiro(1977)]{LS:77}
  Lightman, A. P. \& Shapiro, S. L. 1977,
  ApJ, 211, 244

\bibitem[Makino \& Aarseth(1992)]{Makino:92}
  Makino, J. \& Aarseth, S. J. 1992,
  PASJ, 44, 141

\bibitem[Makino \& Funato(2004)]{Makino:04}
  Makino, J. \& Funato, Y. 2004,
  ApJ, 602, 93

\bibitem[Maoz(1993)]{Maoz:93}
  Maoz, E. 1993,
  MNRAS, 263, 75

\bibitem[Marconi \& Hunt(2003)]{MH:03}
  Marconi, A. \& Hunt, L. K. 2003,
  ApJ, 589, L21

%\bibitem[McClure \& Dunlop(2002)]{McLure:02}
%  McLure, R. J. \& Dunlop, J. S. 2002,
%  \mnras, 331, 795

\bibitem[Merritt(2001)]{Merritt:01}
  Merritt, D. 2001,
  ApJ, 556, 245

\bibitem[Merritt(2004)]{Merritt:04}
  Merritt, D. 2004,
  preprint (astro-ph/0405351)

\bibitem[Merritt(2005)]{Merritt:05}
  Merritt, D. 2005,
  ApJ, 628, 673

\bibitem[Merritt \& Ferrarese(2001)]{MF:01}
  Merritt, D. \& Ferrarese, L. 2001, 
  MNRAS, 320, L30

%\bibitem[Merritt \& Fridman(1996)]{MF:96}
%  Merritt, D. \& Fridman, T. 1996,
%  \apj, 460, 136

\bibitem[Miller(1992)]{Miller:92}
  Miller, R. H. 1992,
  ApJ, 393, 508

\bibitem[Milosavljevi\'c \& Merritt(2001)]{MM:01}
  Milosavljevi\'c, M. \& Merritt, D. 2001,
  \apj, 563, 34

%\bibitem[Milosavljevi\'c et al.(2002)]{Milos:02}
%  Milosavljevi\'c, M., Merritt, D., Rest, A. \& van den Bosch, F. C. 2002,
%  \mnras, 311, L51
%

\bibitem[Preto, Merritt \& Spurzem(2004)]{PMS:04}
  Preto, M., Merritt, D. \& Spurzem, R. 2004,
  preprint (astro-ph/0406324)

\bibitem[Reid et al.(1999)]{Reid:99}
  Reid, M. J., Readhead, A. C. S., Vermeulen, R. C., \& Treuhaft, R. N. 1999,
  ApJ, 524, 816

\bibitem[Reid et al.(2003)]{Reid:03}
  Reid, M. J., Menten, K. M., Genzel, R., Ott, T., Sch\"odel, R.,
  \& Brunthaler, A. 2003,
  Astron. Nachr. 324, No. S1, 3

\bibitem[Salpeter(1955)]{Salpeter:55}
  Salpeter, E. E. 1955,
  ApJ, 121, 161

\bibitem[Sch\"odel et al.(2003)]{Schoedel:03}
  Sch\"odel, R. et al. 2003,
  \apj, 596, 1015

\bibitem[Spitzer(1987)]{Spitzer:87}
  Spitzer, L. 1987,
  Dynamical Evolution of Globular Clusters
  (Princeton: Princeton University Press),
  40

\bibitem[Taga \& Iye(1998)]{Taga:98}
  Taga, M. \& Iye, M. 2998,
  MNRAS, 299, 111

\bibitem[Tremaine et al.(1994)]{Tremaine:94}
  Tremaine, S. et al. 1994,
  AJ, 107, 634

\bibitem[Young(1980)]{Young:80}
  Young, P. J. 1980,
  ApJ, 215, 36

\end{thebibliography}
\end{document}